# Field-Free Multistate Spin-Orbit Torque Switching Enabled by Structurally Engineered Four-Fold Magnetic Anisotropy

Fei Ye(叶飞)[1,†], Sihai Jiao(焦四海)[2,†], Xue Zhang(张雪)[3,†], Chunzheng Wang(王淳正)[2,†], Zhongjie Wang(王忠杰)[2,*], Long Cheng(成龙)[1], Zhifeng Zhu(祝智峰)[3,*], Chunlei Gao(高春雷)[2] & Xiaofang Zhai(翟晓芳)[1,4,*]

[1] School of Physical Science and Technology, ShanghaiTech University, Shanghai 201210, China

[2] State Key Laboratory of Surface Physics, Institute for Nanoelectronic Devices and Quantum Computing, and Department of Physics, Fudan University, Shanghai 200438, China

[3] School of Information Science and Technology, ShanghaiTech University, Shanghai 201210, China

[4] State Key Laboratory of Quantum Functional Materials, School of Physical Science and Technology, ShanghaiTech University, Shanghai 201210, China

†These authors contributed equally to this work.

*Corresponding authors. E-mails:

zhaixf@shanghaitech.edu.cn; zhongjiewang18@fudan.edu.cn; zhuzhf@shanghaitech.edu.cn

**Abstract**

Multiaxial magnetic states offer a route beyond binary spintronic functionalities and enable zero-magnetic-field spin orbit torque (SOT) switching via mirror symmetry breaking, yet remain unexplored. Here, we uncover field-free multistate SOT switching in a $SrIrO_3/SrRuO_3$ bilayer with exceptionally large monoclinic distortion. Pulsed current excitations fully map all twelve deterministic transitions among the four states, through a systematic switching protocol defined by two characteristic current densities. In-situ scanning nitrogen-vacancy (NV) center magnetometry provides direct real-space evidence for previously unobserved in-plane canted states, and spin dynamics simulations uncover a two-step switching pathway, driven by the concerted action of spin torques and the effective anisotropy field within the fourfold anisotropy landscape. Our results demonstrate that precise lattice-symmetry engineering enables the creation and control of multiaxial magnetic states for field-free multistate spintronic devices.

## 1. Introduction

SOT devices that enable nonvolatile magnetization switching via spin current injection are central to next-generation, high speed and energy-efficient spintronics. A typical SOT structure consists of a heavy metal (HM) layer adjacent to a ferromagnetic (FM) layer. When a charge current flows through the HM layer along the *x* direction, the spin-Hall effect (SHE) generates a pure spin current with polarization $\vec{\sigma}$ along the $y$ direction, which is injected into the FM layer.[1-3] The spin current exerts field-like (FL) torque $\vec{\tau}_{\mathrm{FL}} \sim \vec{m} \times \vec{\sigma}$ and damping-like (DL) torque $\vec{\tau}_{\mathrm{DL}} \sim \vec{m} \times (\vec{\sigma} \times \vec{m})$ on the magnetization $\vec{m}$, driving switching dynamics well described by the Landau-Lifshitz-Gilbert (LLG) model.[4] To date, bistable switching between two antiparallel states has advanced significantly, including demonstrations of field-free switching[5-8] and the use of materials with large spin-Hall angles ($\theta_{\mathrm{SH}}$) to minimize switching power.[9,10] Recently, increasing efforts have been devoted to multistate SOT switching as a strategy to fundamentally enhance the storage capacity of binary devices, for example through stacking multiple binary FM layers.[11-13]

Bistate SOT switching in complex oxides has primarily been studied in $SrIrO_3/SrRuO_3$ (SIO/SRO) bilayers, favored by the strong spin–orbit coupling in SIO and the highly tunable magnetic anisotropy (MA) in SRO, which can be perpendicular,[14] canted,[15] and even eight-fold.[16] T. Nan et al. first demonstrated extremely high SOT efficiency in epitaxial metastable SIO films.[17] In SIO/SRO bilayers, both zero-magnetic-field bistate switching and high efficiency up to 0.86 have been achieved via substrate-controlled MA. [7] Furthermore, magnon torques in $SrRuO_3/NiO/SrIrO_3$ heterostructures can efficiently manipulate magnetization with much lower current.[18] These previous progressses make the SIO/SRO heterostructure an excellent platform for exploring possible multistate SOT switching. However, achieving and controlling multiple magnetic states beyond the canted bistate in SIO/SRO bilayers remains a key challenge.

In this work, we fabricate a SIO/SRO bilayer with exceptionally strongly spin canting by growing both layers in low-substrate-temperature and high-oxygen-pressure enviroments. The pulsed current SOT switching reveals the existence of four canted states, which can be switched via two distinct current thresholds with B ≥ 0 and minimal values of $J_{c1} = 6.7\times10^6$ A/cm$^2$ and $J_{c2} = 4.4\times10^6$ A/cm$^2$, comparable to typical bistate devices. Scanning NV-center microscopy visualizes the previously unobserved in-plane (IP) canted states. Spin dynamics simulations reveal that the cooperative action of $\tau_{\mathrm{DL}}$, $\tau_{\mathrm{FL}}$ and the effective anisotropy field $H_{\mathrm{eff}}$ drives a two-step switching pathway via an intermediate state within the four-fold anisotropy landscape. We further demonstrate a four-state read/write protocol enabling all 12 transitions between arbitrary states, with eight via single pulses and four via double pulses. This study opens a new avenue for designing multistate SOT devices based on multiaxial spin-canted magnets.

## 2. Results and Discussion

### 2.1. Exceptionally Strong Monoclinic Distortion and Four-State SOT Switching

The bilayer films have been deposited by reflective high energy electron diffraction (RHEED) assisted pulsed laser deposition (PLD) under a high oxygen enviroment of 10 Pa at a low substrate tempeture of 625 °C for both SIO and SRO. Both SIO and SRO bulks exhibit orthorhombic symmetry and are under compressive strain on STO. According to reference [19], high oxygen pressure growth favors monoclinic structure and spin canting in SRO. In our study, we also find the low substrate temperature also favors the spin canting in SRO. As far as we know, the oxygen pressure and the growth temperature in the current study are both reaching limits in literatures [7,18]. As depicted in Fig. 1, a bulk orthorhombic magnet grown under compressive strain either transforms to tetragonal symmetry (Fig. 1a) with a perpendicular magnetization ($M_{\perp}$), or monoclinic symmetry (Fig. 1b) with tilted magnetization ($M_{\perp}$ and $M_{//}$). Corresponding to Fig. 1a, the SOT switching preserves the mirror symmetry about the *xz* plane with two degenerate antiparallel perpendicular states (Fig. 1c) and external field is required to achieve determinstic switching. In contrast, as shown in Fig. 1d, the monoclinic symmetry and the ±*M* switching are absent of mirror symmetry, which allows zero-field deterministic SOT switching. Indeed, the reciprocal space maps measurements around the pseudocubic $(0\bar{2}4)$, (204), (024) and $(\bar{2}04)$ peaks are shown in Fig. 1e, which demonstrates an overall *c*-axis canting anlge of 0.26° ± 0.01°, which is larger than the largest canting angle reported in ref. [19]. Meanwhile, the bilayer film exhibits high-quality FM conduction, with a well-defined FM transition at 142 K (Fig. 1f), comparable to the optimal values observed in bulk single crystals and thick films.[15] Furthermore, we observed the IP saturation moment under a small field to be nearly 40% of the out-of-plane (OP) moment (Fig. 1g), corresponding to 26° canting angle of the easy axis. Consistent results were obtained from the zero-field magnetization-temperature (MT) data (Fig. S1). To further elucidate this distinct magnetic anisotropy configuration, we provided angle-dependent magnetoresistance (AMR) data in the *yz* plane in Fig. S2. Notably, anomalous extrema emerge near 90° and 270°, exhibiting a distorted four-fold symmetry that demonstrates the presence of an in-plane easy axis. Concurrently, the magnetoresistance minima occur at 25° and 205°. This corresponds to a 25° tilt from the out-of-plane direction, which is highly consistent with MT measurement results. Fig. 1h demonstrates that the ratio is much higher than those in previous SIO/SRO SOT devices.[7, 20] Additonal details of the fabrication, structural characterization, magneto-transport and magnetization measurements are provided in the Experimental Section, Supplementary Information Fig. S3 and Fig. S4.

The schematic of the SIO/SRO bilayer lattice, the Hall-bar device geometry and the SOT switching examples ($\mathrm{OP_c^+} \rightarrow \mathrm{IP_c^+}$ and $\mathrm{IP_c^+} \rightarrow \mathrm{OP_c^-}$) are shown in Fig. 2a, b. Here, the '+/−' signs denote positive/negative IP or OP magnetization directions, and the subscript 'c' indicates canting. The pulsed-current switching and ac-current detecting sequence is shown in Fig. S5. The representative SOT switching loops are shown in Fig. 2c, which have been measured at 60 K under external magnetic fields of $H_\mathrm{x} = 0$, 100, 300 and 1000 Oe. Unlike bistate devices, four distinct anomalous Hall resistance ($R_H$) plateaus are observed in all switching loops: the top (defined as $\mathrm{OP_c^+}$) and bottom (defined as $\mathrm{OP_c^-}$) plateaus appear at small $|J|$, while the left (defined as $\mathrm{IP_c^-}$) and right (defined as $\mathrm{IP_c^+}$) plateaus arise at large $|J|$, as exemplified by the 300 Oe loop in Fig. 2c. Between these plateaus, $R_H$ exhibits clear transitions: from $\mathrm{OP_c^\pm}$ to $\mathrm{IP_c^\pm}$ when $|J|$ exceeds $J_{c1}$, and from $\mathrm{IP_c^\pm}$ to $\mathrm{OP_c^\mp}$ when $|J|$ decreases to $J_{c2}$. Similar plateau structures and transition behaviors are also observed in loops measured under negative $H_x$ (Fig. S6). Fig. 2d summarizes the dependence of $J_{c1}$ and $J_{c2}$ on $H_x$, both decreasing with the increasing field when $H_x \leq$ 100 Oe. But they behave differently when $H_x >$ 100 Oe: $J_{c1}$ continuously decreases with increasing $H_x$ reaching $7.3\times10^6$ A/cm$^2$ when $H_x$ = 1000 Oe, while $J_{c2}$ remains constant at $5.5\times10^6$ A/cm$^2$. The distinct field dependences of $J_{c1}$ and $J_{c2}$ indicate that the two types of switching processes are governed by different underlying mechanisms.

We further investigate the temperature dependence of the four-state SOT switching. Fig. S7 shows the switching loops measured between 2 K and 80 K under a constant $H_x$ = 1000 Oe, which consistently exhibit robust four-state switching behaviors. The switching loops at higher temperatures are presented in Figure S8. At 100 K, the magnetization switching degrades significantly. Therefore, the operating temperature window for well-defined four-state behavior spans from 2 K to 90 K. The extracted $J_{c1}$ and $J_{c2}$ are shown in Fig. 2e, which progressively decrease with increasing temperature, likely due to the reduction in saturation magnetization at elevated temperatures.[21] As $T$ increases from 2 K to 80 K, $J_{c1}$ and $J_{c2}$ decrease by about 35%, reaching minima of $6.7\times10^6$ A/cm$^2$ and $4.4\times10^6$ A/cm$^2$, respectively. The corresponding power dissipations are estimated using $P_{\mathrm{sw}} = \rho_{xx} J_c^2$, where $\rho_{xx}$ is the longitudinal resistivity, yielding $P_{\mathrm{sw1}} = 1.03\times10^{16}$ W/m$^3$ and $P_{\mathrm{sw2}} = 4.6\times10^{15}$ W/m$^3$. We further benchmark these $P_{\mathrm{sw1,2}}$ values, together with the reported spin-Hall angle of SIO ($\theta_{\mathrm{SH}}$~0.5),[17] against state-of-the-art bistate SOT devices across various material platforms. As summarized in Fig. 2f, the four-state SOT device achieves energy efficiencies comparable to many advanced bistate devices,[1,2,7,22-

[32] suggesting considerable potential for future optimization. For comparison, the pure SRO device exhibits no switching loops, confirming that spin current is primarily contributed by SIO (Fig. S9).

**2.2. In-situ NV-center Magnetometry Imaging and Spin Dynamics Simulations**

To reveal the microscopic switching process, we perform real-space magnetic imaging using in-situ scanning NV-center magnetometry. The in-situ measurement setup and the single spin direction of the NV-center (NV axis) are schematically illustrated in Fig. 3a, b. The NV-axis resides in the *yz* plane, tilted 54.7° from the *z*-axis. Current pulses and an external field of 300 Oe are applied along the *x*-axis immediately before NV measurements. More measurement details are described in the Experimental Section. Micrometer-scale domain structures corresponding to $\mathrm{IP_c^+}$ and $\mathrm{IP_c^-}$ states, obtained after +20 mA and −20 mA pulses respectively, are presented in Fig. 3c and d. A sharp contrast is observed in these magnetic textures, with most regions exhibiting a clear reversal of the stray field (indicated by guidelines).This shows that the domain configuration is preserved while the magnetization within each domain is inverted, implying that domains in the $\mathrm{IP_c^\pm}$ states can be collectively reversed. Notably, the NV sensing axis is approximately aligned with the $\mathrm{IP_c^\pm}$ magnetization, as schematically depicted in Fig. 3b, which limits the projection of the latter onto the detection axis and may hinder clear visualization of $\mathrm{OP_c^\pm}$ state. Consequently, the strong contrast observed between the $\mathrm{IP_c^\pm}$ maps implies a substantial projection of their magnetization along the NV axis (*y*-direction), agreeing to SQUID measurements shown in Fig. 1g.

Next, we propose possible mechanisms for the four-state switching. Since the damping-like torque $\tau_{\mathrm{DL}}$ consistently drives the magnetization $\vec{m}$ towards the injected spin-current polarization direction ($\pm\vec{y}$), $\tau_{\mathrm{DL}}$ is likely responsible for the switching from OP to IP, as shown in Fig. 3e. In comparison, the field-like torque $\tau_{\mathrm{FL}}$ and external torque $\tau_{\mathrm{ext}}$ are precessional in nature, with dynamically evolving directions. In the IP state (Fig. 3f), $\tau_{\mathrm{FL}}$ and $\tau_{\mathrm{ext}}$ approximately act in the same direction when $m_x > 0$, but oppose each other when $m_x < 0$. When the current amplitude falls below a critical threshold, the magnitude of $\tau_{DL}$ declines, rendering it insufficient to maintain alignment along the $IP_c^{\pm}$ easy axis. Under these conditions, the combined effects of $\tau_{FL}$ and $\tau_{ext}$ may drive the precessional motion of $\vec{m}$ away from the $IP_c^{\pm}$ easy axis, eventually allowing relaxation into the $OP_c^{\pm}$ states.

To confirm the above proposed mechanism, we perform the spin dynamics simulations[33] by solving the Landau-Lifshitz-Gilbert-Slonczewski (LLGS) equation: $\frac{\partial\vec{m}}{\partial t} = -\gamma\vec{m} \times (\vec{H}_{eff} + \vec{H}_x) + \alpha\vec{m} \times \frac{\partial\vec{m}}{\partial t} - \gamma B_D\vec{m} \times (\vec{m} \times \vec{\sigma}) + \gamma B_F\vec{m} \times \vec{\sigma}$. The four terms on the right-hand side represent precession, Gilbert damping, DL torque and FL torque, respectively. Here, $\gamma$ is the gyromagnetic ratio and the effective field is $\vec{H}_{eff} = -\frac{\partial\mathcal{H}}{\partial\vec{m}}$. $B_D$ and $B_F$ denote the magnitudes of the DL and FL torques, respectively. The Hamiltonian is $\mathcal{H} = K_u sin\theta^2 - K_y sin\theta^2 sin\varphi^2 - K_x sin\theta^2 cos\varphi^2 - K_{4u}\cos(4\theta)$, where $K_u$, $K_y$ and $K_x$ are the uniaxial anisotropy constants along the *z*, *y*, and *x* axes, and $K_{4u}$ represents the fourfold anisotropy energy,[34] indispensable for the four-state switching. $\theta$ and $\varphi$ are the polar and azimuthal angles in the spherical coordinate. In accordance with the experiment setup, a current $J$ along $+\vec{x}$ induces spin polarization $\vec{\sigma}$ along $-\vec{y}$. Magnitudes of $B_D$, $K_u$, $K_y$ and $K_x$ are based on experiments, while $B_F = \beta B_D$ is optimized to reproduce the four-state switching loop in Fig. 4b, consistent to the experiment. More parameter optimization details are provided in Experimental Section and Supplementary Information (Fig. S10 and Fig. S11). The energy diagram in Fig. 4a derived from the Hamiltonian reveals the four ground states corresponding to $OP_c^{\pm}$ and $IP_c^{\pm}$. The two pairs of mutually orthogonal easy axes are rotated clockwise by ~20° from the *z*-axis within the *y*-*z* plane, in good agreement to the SQUID and NV-center measurements. As shown in Fig. 4b, the $OP_c^{\pm}$ and $IP_c^{\pm}$ states exhibit magnetization components of $m_z = \pm 0.94$ and $\mp 0.34$, respectively.

We plot the simulated switching trajectories in Fig. 4c, d, which reveal a distinct a two-step switching process. Taking the switching from $\mathrm{OP_c^+}$ to $\mathrm{IP_c^+}$ (Fig. 4c) as an example, the black arrow marks the initial stable state ($\vec{m}_{\mathrm{init}}$). As previously described, upon current $J$ excitation, $\vec{m}_{\mathrm{init}}$ begins to precess and flip under the influence of $\vec{\tau}_{\mathrm{FL}}$and $\vec{\tau}_{\mathrm{DL}}$. However, it stabilizes at an intermediate stable state ($\vec{m}_{\mathrm{mid}}$), whose direction is mainly determined by the balance between $\vec{\tau}_{\mathrm{FL}}$ and $\vec{\tau}_{\mathrm{H_{eff}}}$ (torque induced by $\vec{H}_{\mathrm{eff}}$). If $J$ is below the threshold $J_{\mathrm{c1}}$, $\vec{m}_{\mathrm{mid}}$ relaxes back to $\vec{m}_{\mathrm{init}}$ once the pulse is removed (see the time evolution of $m_x$, $m_y$ and $m_z$ in Fig. 4g). But when $J \geq J_{\mathrm{c1}}$, $\vec{\tau}_{\mathrm{FL}}$ (green arrow in Fig. 4c) that exerts a $-\hat{z}$ direction torque, becomes strong enough to reduce the $m_z$ component of $\vec{m}_{\mathrm{mid}}$ below a critical level. This is the key to the successful transition to $\mathrm{IP_c^+}$. After the pulse is turned off, $\vec{\tau}_{\mathrm{H_{eff}}}$ guides the moment toward the $-\hat{z}$ direction and settles it into the stable $\mathrm{IP_c^+}$ state, i.e., $\vec{m}_{\mathrm{final}}$. The corresponding time evolutions of $m_x$, $m_y$ and $m_z$ for $J = J_{\mathrm{c1}}$ are shown in Fig. 4e, where the intermediate state remains metastable for most of the 20 ns pulse duration, except the initial ~2 ns transition period.

However, the mechanism underlying the $\mathrm{IP_c^+}$ to $\mathrm{OP_c^-}$ transition differs markedly from the above case. As shown in Fig. 4d, the blue arrow represents the torque from the in-plane field $\vec{H}_{\mathrm{x}}$ ($\vec{\tau}_{\mathrm{H_x}}$). In the above $\mathrm{OP_c^+} \rightarrow \mathrm{IP_c^+}$ transition, $\vec{\tau}_{\mathrm{H_x}}$ and $\vec{\tau}_{\mathrm{H_{eff}}}$ act in the same direction and thus are not separated. In contrast, for the $\mathrm{IP_c^+} \rightarrow \mathrm{OP_c^-}$ transition, $\vec{\tau}_{\mathrm{H_x}}$ is no longer aligned with $\vec{\tau}_{\mathrm{H_{eff}}}$, but instead opposes the $z$-component FL torque ($\vec{\tau}_{\mathrm{FL,z}}$). Specifically, $\vec{\tau}_{\mathrm{FL,z}}$ pushes the moment to the $+\hat{z}$ direction ($\vec{m}_{\mathrm{init}}\times\vec{\sigma} \propto (-\hat{x}) \times (-\hat{y}) = \hat{z}$), whereas $\vec{\tau}_{\mathrm{H_x}}$ acts oppositely ($-\vec{m}_{\mathrm{init}} \times \vec{H}_{\mathrm{x}} \propto -(+\hat{y}) \times (-\hat{x}) = -\hat{z}$). When $J > J_{\mathrm{c2}}$, $\vec{\tau}_{\mathrm{FL,z}}$ dominates and displaces $\vec{m}_{\mathrm{init}}$ away from the $\mathrm{OP_c^-}$ anisotropy axis. Upon removing $J$, the $\mathrm{IP_c^+}$ anisotropy prevails, causing $\vec{m}_{\mathrm{mid}}$ to relax back to $\vec{m}_{\mathrm{init}}$ (see the time revolution in Fig. 4h). As $J$ is reduced from a large value to $J_{\mathrm{c2}}$, $\vec{\tau}_{\mathrm{H_x}}$ overcomes $\vec{\tau}_{\mathrm{FL,z}}$ and $\vec{m}_{\mathrm{mid}}$ moves closer to the $\mathrm{OP_c^-}$ anisotropy axis (Fig. 4f). Therefore, once $J$ is turned off, $\vec{\tau}_{H_x}$ together with the $OP_c^-$ anisotropy drives the system into the $OP_c^-$ state (Fig. 4d). When $J$ is further reduced, the magnetization remains trapped in the $IP_c^+$ state. Furthermore, as shown in Fig. S12 (Supplementary Information), the switching trajectories and time-dependence for the $OP_c^- \rightarrow IP_c^-$ and $IP_c^- \rightarrow OP_c^+$ transitions are symmetric to those of $OP_c^+ \rightarrow IP_c^+$ and $IP_c^+ \rightarrow OP_c^-$, respectively, reflecting their shared underlying mechanisms.

### 2.3. Mapping Four-State Switchings and Programmable Writing/Reading

We thoroughly map all twelve transitions that a four-state device can maximally have. As shown in Fig. 5a, b, the four states are encoded as '0' ($OP_c^+$), '1' ($IP_c^-$), '2' ($IP_c^+$) and '3' ($OP_c^-$) to facilitate programming. The gray trajectories exhibit four transitions described previously, where the initial state is '0'. The newly explored trajectories are marked in red and blue, where the initial state is '2' (Fig. 5a) or '1' (Fig. 5b). In the red trajectory of Fig. 5a, after 'writing' the device into state '2' with $J \geq J_{c1}$, $J$ is turned off and then gradually increased from zero. The state '2' switches to '3' at $J = J_{c2}$, but '3' reverts to '2' when $J \geq J_{c1}$. Along the blue trajectory (negative $J$), state '2' is switched to '1' when $J \leq -J_{c1}$. The latter two processes ('3'→'2', '2' →'1') represent newly observed transitions. Similarly, in Fig. 5b, another two new transitions of '0'→'1' and '1' →'2' are revealed. All eight single-pulse transitions are summarized in Fig. 5c, where orange and blue arrows indicate thresholds of $|J| \geq J_{c1}$ and $|J| = J_{c2}$, respectively. Solid and dash arrows denote positive and negative current. The remaining four transitions ('0'→'3', '2'→'0', '3'→'0', and '1'→'3'), all terminating in the $OP_c^{\pm}$ states, require double-pulse excitation, making them particularly robust and suitable for fault-tolerant data writing. The corresponding spin dynamics simulations are shown in Fig. S13. Note such complex switching trajectory has been achieved in the absence of external field (see Fig. 5f).

We evaluate the stability of the previously unexplored '1' and '2' states ($IP_c^{\pm}$). As shown in the left panel of Fig. 5d, the device is initialized into state '2' after completing the $R_H$ vs. $J$ loop with a final pulse of $J > J_{c1}$. Subsequently, $R_H$ is continuously monitored using a low-amplitude AC current (Fig. 5d right panel). The measured $R_{\mathrm{H}}$ remains highly stable over an extended period of about three hours, exhibiting a standard deviation of ±1.84% over 2096 readings under a 300 Oe in-plane field, and ±3.4% over 2860 readings at zero field.

Finally, we demonstrate the programmable writing and reading of the four-state SOT device both with (Fig. 5e) and without (Fig. 5g) an applied magnetic field. Based on the established switching protocol, a positive pulse with $J \geq J_{\mathrm{c1}}$ sets the device into state '2' and a negative pulse with $J \leq -J_{\mathrm{c1}}$ sets state '1', both achievable from any initial state. To set the '3' state from an arbitrary initial state, a double-pulse excitation sequence of ($J \geq J_{\mathrm{c1}}$ and $J = J_{\mathrm{c2}}$) is required. Similarly, a sequence of ($J \leq -J_{\mathrm{c1}}$ and $J = -J_{\mathrm{c2}}$) always set the state '0'. As shown in Fig. 5e, g, the programmed writing follows the sequence '2'→'1', '3'→'0' →.... →'3'→'1', each followed by consecutive readings. Indeed, each target state is well written and maintained throughout the entire reading period.

## 3. Conclusions

By fabricating SRO/SIO bilayer films with strongly canted FM via high-oxygen-pressure and low-substrate-temperature PLD growth, we realize a zero-field multistate SOT switching consisting of four stable yet electrically distinguishable spin-canting states, $\mathrm{IP_c^{\pm}}$ and $\mathrm{OP_c^{\pm}}$. Among these states, eight distinct transitions are demonstrated by single-pulse current excitations and four by double-pulse excitations, all governed by two critical current densities that define the complete four-state switching protocol. NV-center magnetometry and spin dynamics simulations reveal that the effective anisotropy field $H_{\mathrm{eff}}$ of the four-canted configuration plays a pivotal role, acting in concert with spin torques to direct each switching pathway. Furthermore, we demonstrate all 12 switching paths and programmable switching both with and without applied magnetic field. The zero-field multistate SOT switching principle revealed here is broadly applicable to other multiaxial magnets across diverse material families.

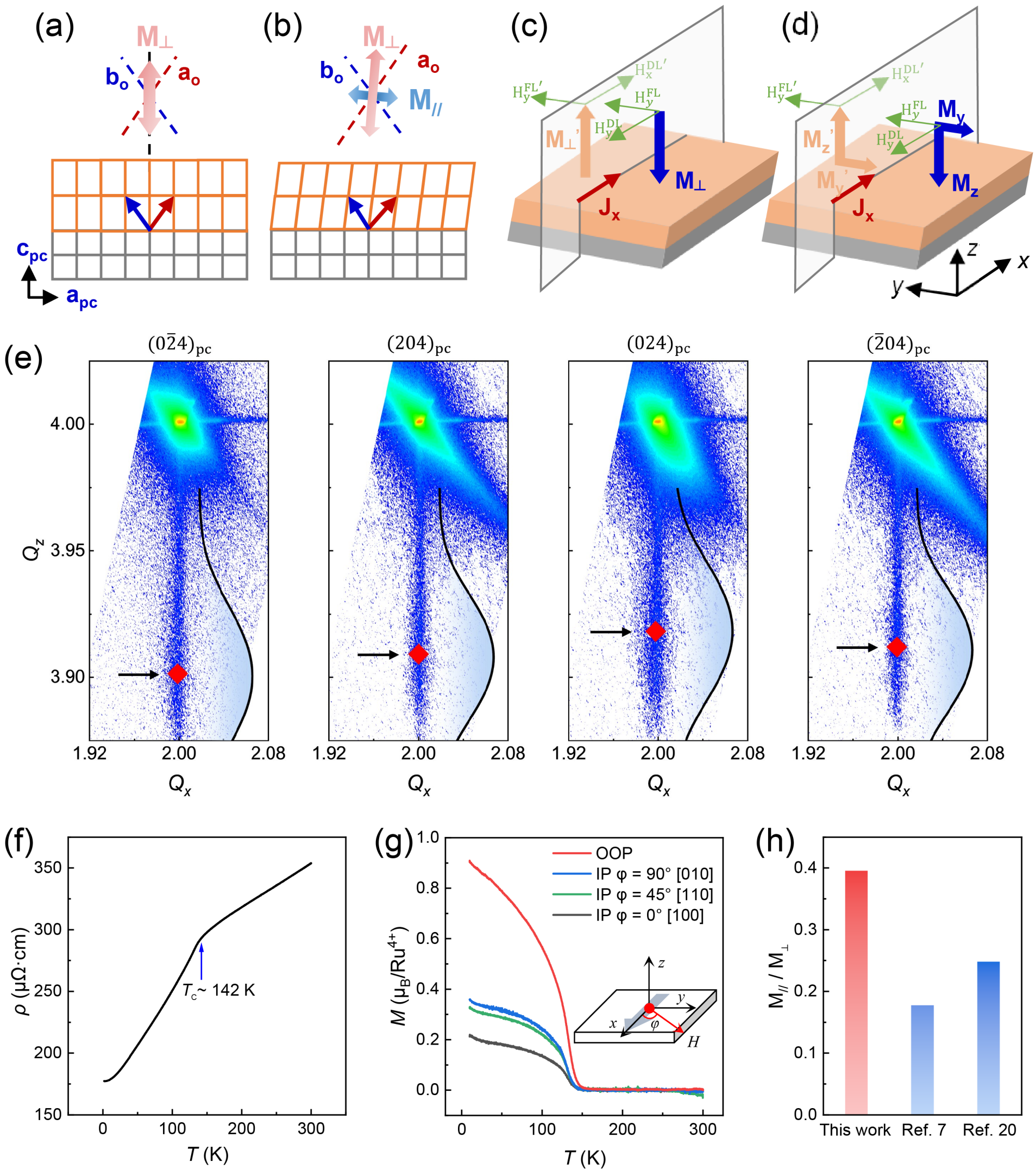

**Fig. 1.** Structure symmetry engineering and prominant spin canting. (a) Tetragonal symmetry under compressive strain and perpendicular magnetic anisotropy; (b) Monoclinic symmetry under compressive strain and canted magnetic anisotropy; (c) Preserved mirror symmetry about *xz* plane with two symmetry-equivalent antiparallel magnetic states, corresponding to Fig. (a); (d) Broken mirror symmetry about *xz* plane and symmetry-inequivalent ±M states, corresponding to Fig. (b); (e) Reciprocal space maps around $(0\bar{2}4)_{pc}$, $(204)_{pc}$, $(024)_{pc}$ and $(\bar{2}04)_{pc}$ diffractions of the STO substrate and SIO/SRO bilayer film; The subscript 'pc' stands for pseudocubic. The film diffraction peaks are underneath the bright substrate diffraction peaks. (f) Temperature dependence of resistivity of the bilayer film. (g) Magnetization versus temperature curves measured with an applied field of 1000 Oe. Inset: schematic diagram of the measurement geometry. (h) Comparison of $M_{\parallel}$ / $M_{\perp}$ ratios between this work and those in literature.[7,20]

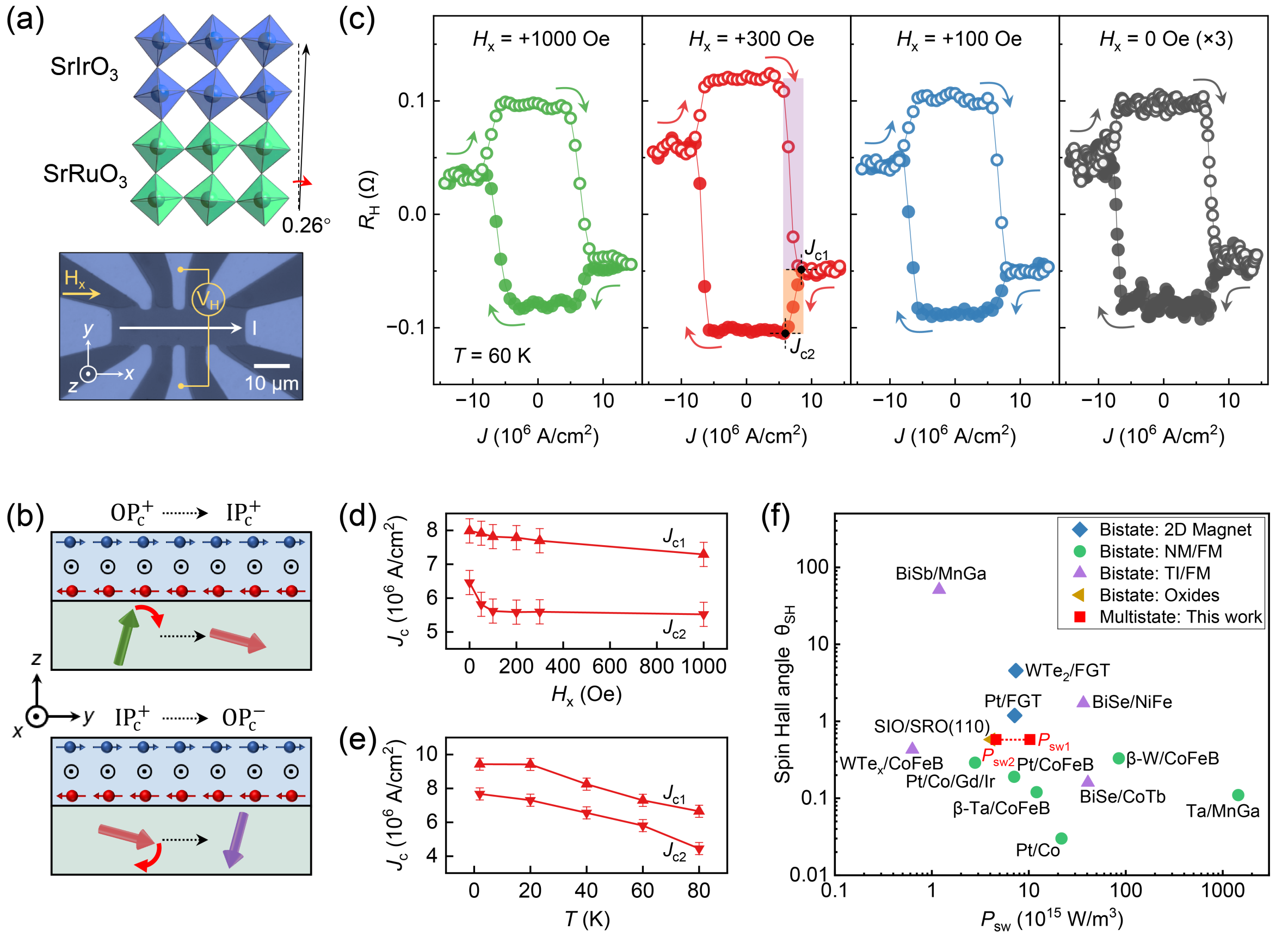

**Fig. 2.** Current-induced magnetization switching in a four-state SOT device. (a) Schematic illustration of the bilayer structure and optical image of the Hall-bar device and measurement schematics. (b) Cartoons of spin current generation and switching of the FM moment from $OP_c^+$ to $IP_c^+$ and from $IP_c^+$ to the $OP_c^-$. (c) Magnetization switching loops at 60 K. Arrows indicate the current sweeping directions. (d) The critical current densities of $J_{c1}$ and $J_{c2}$ as functions of in-plane field $H_x$ at 60 K, and (e) as functions of temperature with fixed $H_x$ = 1000 Oe. (f) Spin-Hall angle $\theta_{SH}$ and the switching power $P_{sw}$ of advanced bistate SOT devices [Ref. 1, 2, 7, 22-32] and the current four-state device (data taken at 80 K). The two red points denote the $P_{sw1}$ (right) and $P_{sw2}$ (left), respectively.

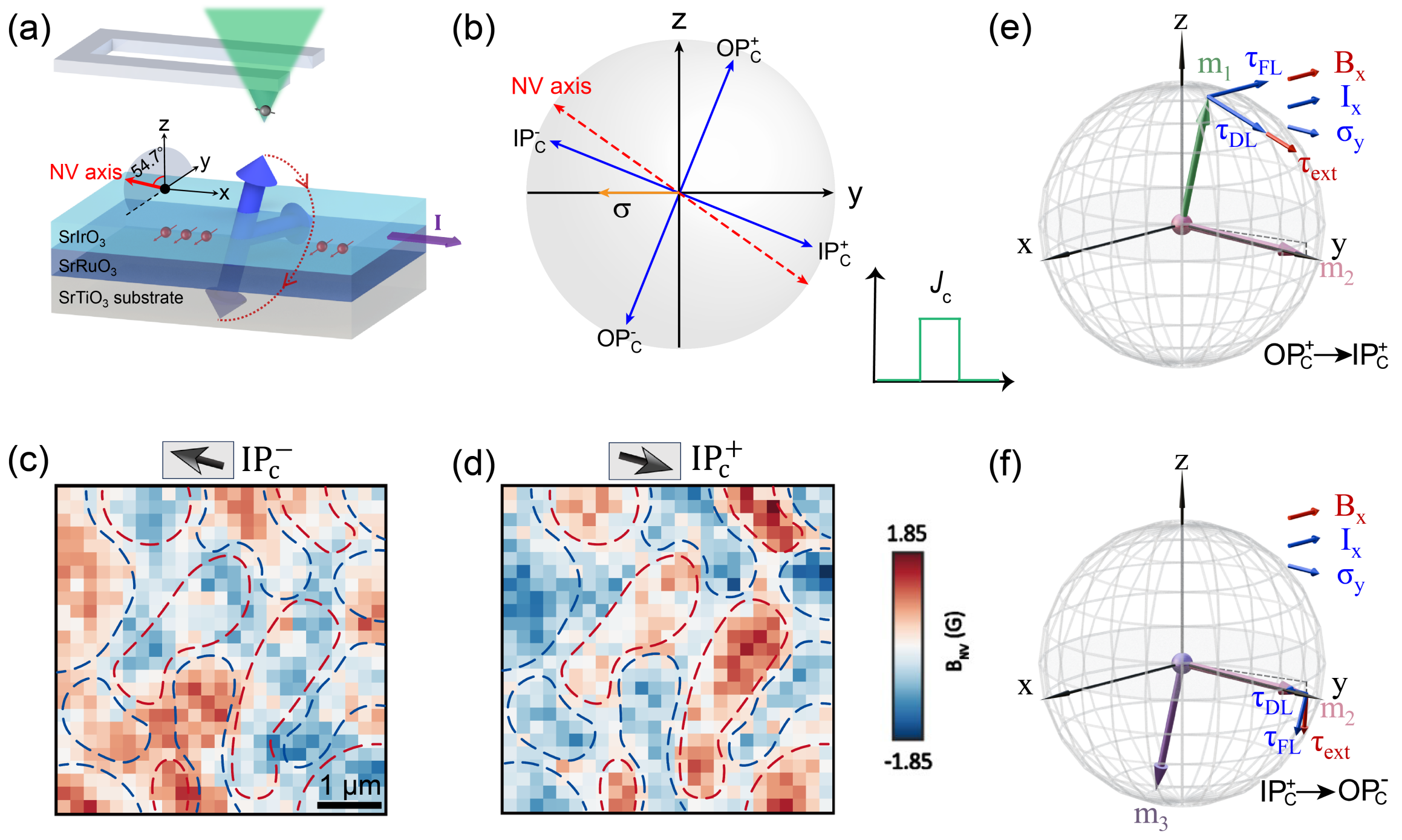

**Fig. 3.** In-situ NV-center magnetometry imaging. (a) Schematic of the scanning NV-center magnetometer. The variation of the magnetization (blue arrow) in SRO can be detected by the NV-center through changes in the magnetic field. (b) Configurations of relative orientations between the NV-axis and magnetizations. (c), (d) Magnetic textures corresponding to $IP_c^-$ (c) and $IP_c^+$ (d) revealed by NV-center; The red and blue dashed lines delineate the contours of magnetic domains. (e), (f) Multiple magnetic torques for $OP_c^+$ (e) and $IP_c^+$ (f) states.

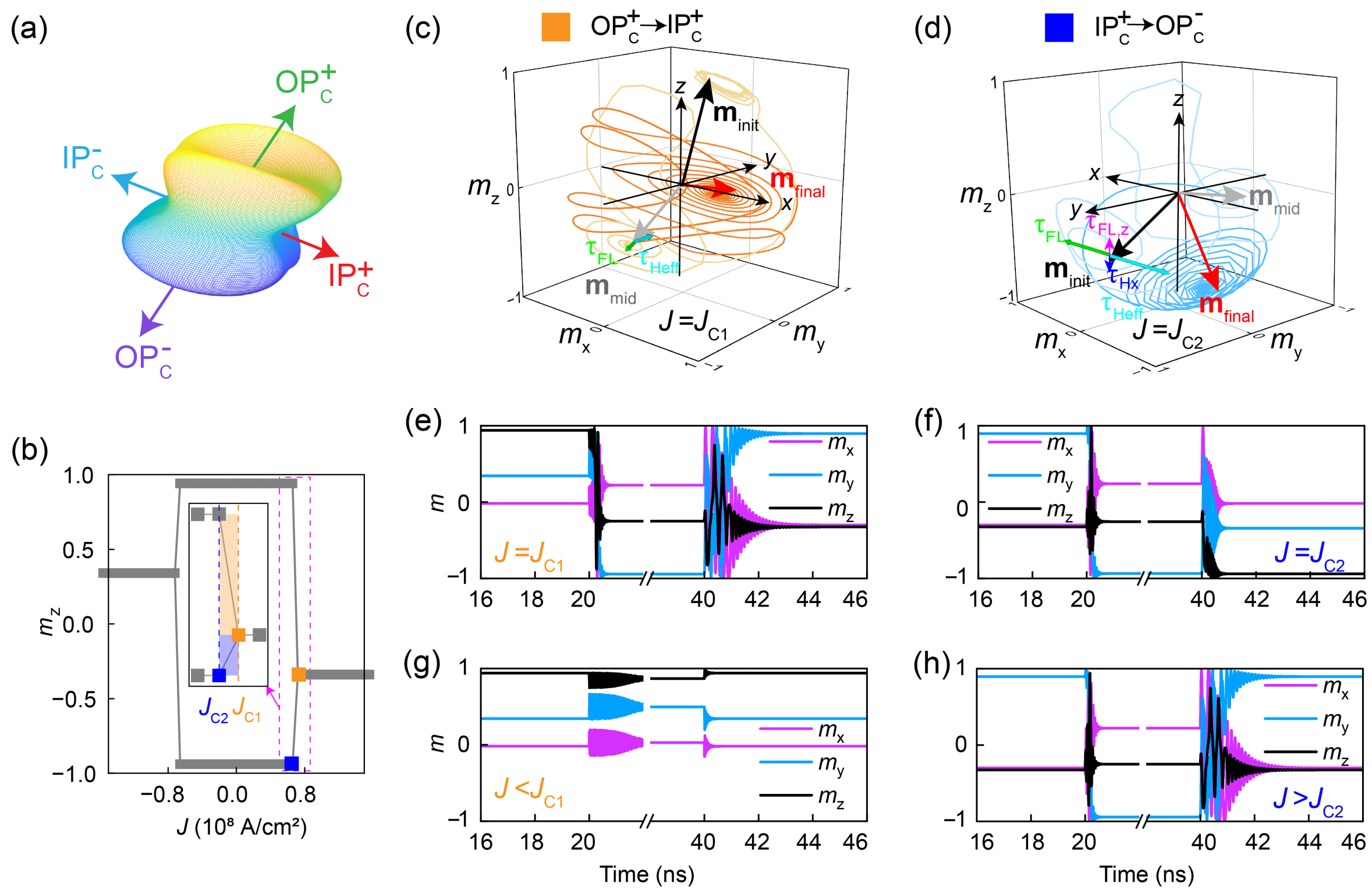

**Fig. 4.** The spin dynamics simulations of the four-state switching loop. (a) The energy diagram of the system. (b) The $m_z$ vs. $J$ loop under $H_x$ = –300 Oe and $\beta$ = 3. The switching trajectory for (c) $OP_c^+$ to $IP_c^+$, (d) $IP_c^+$ to $OP_c^-$. The black, gray and red arrows represent the $\vec{m}_{init}$, $\vec{m}_{mid}$, $\vec{m}_{final}$, respectively. The time evolution of ($m_x$, $m_y$, $m_z$) when (e) $J = J_{c1}$, (f) $J = J_{c2}$, (g) $J < J_{c1}$, and (h) $J > J_{c2}$.

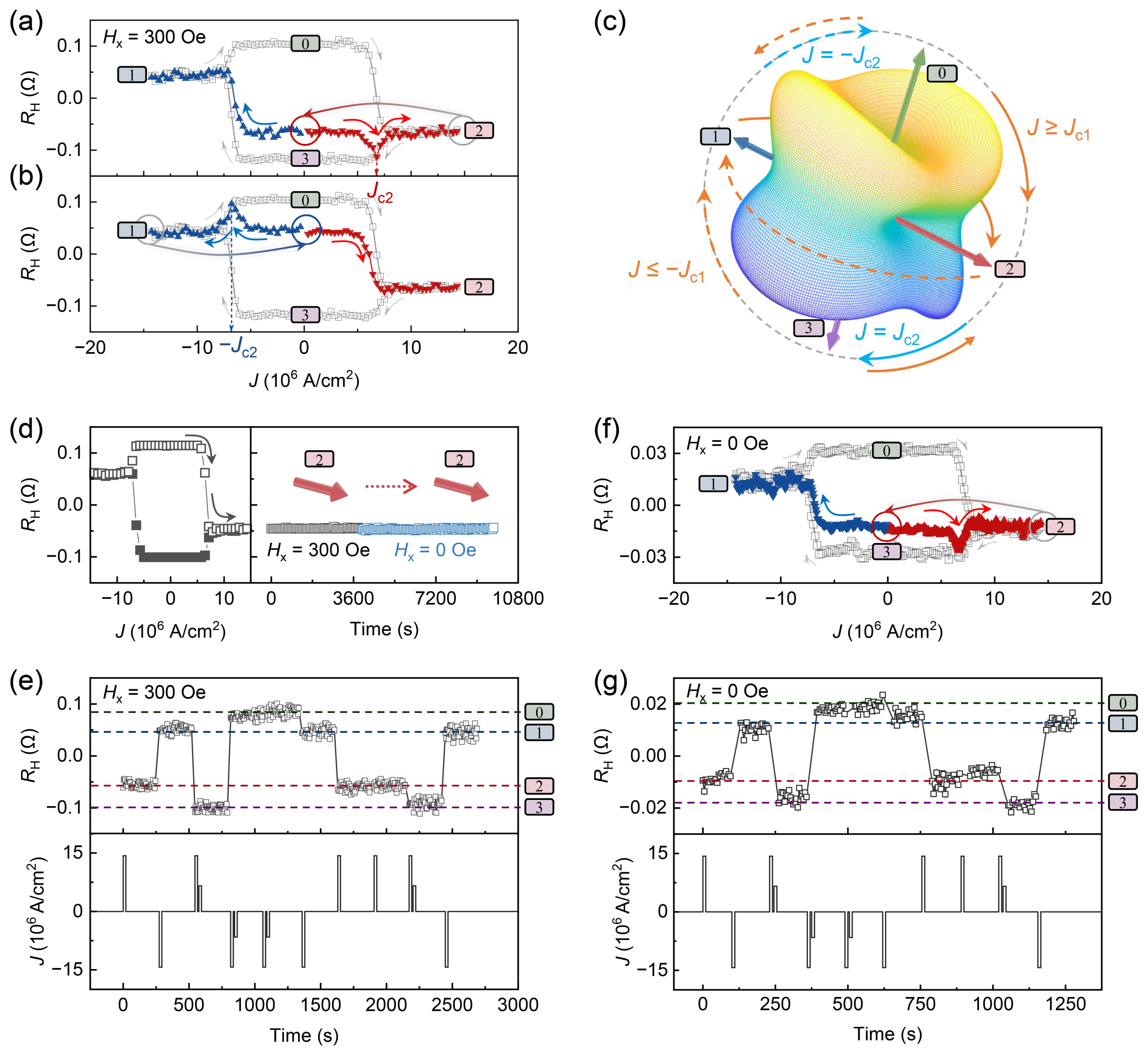

**Fig. 5.** Mapping transition paths, stability verification and programmable writing/reading. (a) The blue and red trajectories represent transition paths starting from state ‘2’ (IP$_c^+$), and (b) from state ‘1’ (IP$_c^-$), both under $H_x$ = +300 Oe at 60 K. The gray color trajectory represents the loop starting from state ‘0’, which has been discussed in Fig. 2-4. (c) Schematic illustrations of the eight single-pulse switching paths. The orange and blue arrows represent $J \geq J_{c1}$ and $J = J_{c2}$, respectively. Solid (dashed) arrows indicate positive (negative) $J$. (d) Temporal stability of the state ‘2’ after being prepared by $J > J_{c1}$. (e) Programmed writing and reading of the four-state SOT device under $H_x$ = +300 Oe; (f) Transition paths starting from state ‘2’ (IP$_c^+$) under zero field; (g) Programmed writing and reading under zero field.

## 4. Experimental Section

*Sample growth and device fabrication*: The SRO (6 nm) and SIO (8 nm) bilayer film was epitaxially grown on a 5 × 5 $mm^2$ $TiO_2$-terminated $SrTiO_3$ (001) substrate employing in-situ reflection high-energy electron diffraction (RHEED) assisted pulsed laser deposition (PLD). During deposition, the substrate temperature was 625 °C and the oxygen pressure was 10 Pa. The laser energy fluence was 1.35 J/$cm^2$ and 1.25 J/$cm^2$ for depositing SRO and SIO, respectively. After deposition, samples were cooled down to room temperature at a rate of 20 °C/min under the same oxygen pressure of 10 Pa. Hall bar patterns with a channel width of 10 μm were transferred to the bilayer films using laser writing and Ar ion milling. Finally, the Ti (10 nm)/Au (60 nm) electrodes were deposited utilizing a magnetron sputtering system.

*Electrical transport and magnetization measurements*: The magneto-transport measurements were performed in a $He_4$ refrigerator (Oxford Teslatron PT system). For current-induced switching measurements, the current pulse sequences were custom-programmed, and each d.c. pulse was sequentially generated by a Keithley 6221 and applied to the current path of the Hall bar. Each current pulse was precisely set to 30 μs by configuring the current source in square-wave mode with a specific duty cycle. After each pulse and a sufficiently long cooling time (10 s), a small a.c. excitation (10 μA) was applied to measure the $R_H$. The a.c. voltage measurements were performed using Stanford Research Systems SR830. The SOT current pulse excitation and measurement sequences are illustrated in Fig. S5. The magnetic properties were measured using a Quantum Design Magnetic Properties Measurement System.

*Scanning NV-center microscopy set-up and measurement principle*: All NV-based magnetometry measurements were performed using a custom-built ultra-high-vacuum, NV-center scanning microscope developed in collaboration with CIQTEK. The system operates over a temperature range of 4-300 K and is equipped with a three-axis vector magnet providing fields of 300, 300 and 1500 Oe along the *x*, *y*, and *z* directions, respectively.

During the NV measurements, the sample was polarized by current pulses from -20 mA to +20 mA, following the switching loop and under an in-plane field of 300 Oe. After the current-pulse switching, the 300 Oe magnetic field was removed while a $B_{\mathrm{bias}} = 20$ Oe field was applied along the NV axis during the magnetic imaging. We measured the magnetic field above the sample based on optically detected magnetic resonance (ODMR). The negatively charged NV center exhibits a spin-triplet ground state. With an external magnetic field $B_{\parallel}$ applied along the NV axis, the spin states $|m_{\mathrm{s}} = \pm 1\rangle$ exhibit Zeeman splitting of $f_{\pm 1} = D_{\mathrm{s}} \pm \gamma_{\mathrm{e}} B_{\parallel}$ from $|m_{\mathrm{s}} = 0\rangle$, where $D_s = 2.87$ GHz is the zero field splitting, and $\gamma_{\mathrm{e}} = 28$ GHzT$^{-1}$is the electronic spin gyromagnetic ratio. By sweeping the microwave frequency and monitoring the fluorescence, both frequency $f_{+1}$ and $f_{-1}$ are tracked and the $B_{\parallel}$ above the sample can be determined by $B_{\parallel} = \frac{|f_{+1} - f_{-1}|}{2\gamma_{\mathrm{e}}} - B_{\mathrm{bias}}$.

*Spin dynamics simulations*: The magnetization dynamics were simulated by solving the LLG equation with SOT terms. The magnetic anisotropy energies were set as $K_{\mathrm{u}} = 3.25\times10^5$ J/m$^3$, $K_{\mathrm{y}} = 1.625\times10^5$ J/m$^3$, and $K_{\mathrm{x}} = 0.975\times10^5$ J/m$^3$, corresponding to the uniaxial anisotropy constants along the *z*, *y*, and *x* axes, respectively. These values are chosen to reflect the magnetization preference along the *x*-, *y*-, *z*-axis in the experimental magnetization data in Fig. 1g. In addition, a fourfold anisotropy term $K_{4\mathrm{u}} = 0.975\times10^5$ J/m$^3$ was included.[34] $B_{\mathrm{D}} = \frac{\hbar}{2}\frac{J_{\mathrm{c}}\theta_{\mathrm{SH}}}{M_{\mathrm{s}}et}$ and $B_{\mathrm{F}}$ represent the magnitude of DL and FL torques, respectively, where $\theta_{\mathrm{SH}} = 0.5$ is the spin-Hall angle of SIO. $M_{\mathrm{s}}$ and $t$ are the saturation magnetization and the FM layer thickness, in agree with the experimental values.

**Acknowledgements**

We thank Kaiyou Wang and Xuepeng Qiu for valuable discussions. This work was supported by the National Key Research and Development Program of China (Grant No.

2023YFA1406301, 2022YFA1403000), the National Natural Science Foundation of China (Grant No. 62574134), the Natural Science Foundation of Shanghai (Grant No. 25ZR1402375), and the China Postdoctoral Science Foundation (Grant No. KLH1512149). The simulations conducted in this work is supported by HPC Platform and SIST Computing Platform at ShanghaiTech University. The research used resources from Analytical Instrumentation Center (#SPST-AIC10112914) and Soft Matter Nanofab (SMN180827) in ShanghaiTech University. XF.Z. and Z.Z. acknowledge start-up grants of ShanghaiTech University.

**Competing Interests**

The authors declare no competing interests.

**Author Contributions**

XF.Z. and Z.W. conceived the idea and designed experiments. XF.Z. supervised the project. F.Y. fabricated samples, performed the electric transport and magnetization measurements with help from L.C.. S.J., C.W., and Z.W. performed the NV-center experiments. X.Z. and Z.Z. performed the spin dynamics simulations. All authors discussed the results in the paper. Z.W., F.Y., X.Z., and C.W. wrote the initial paper, which was revised by XF.Z., Z.Z. and C.G., and approved by all authors.

**Data Availability Statement**

The data that support the findings of this study are available from the corresponding author upon reasonable request.